\documentclass[journal]{IEEEtran}
\usepackage{amsmath,amsfonts,graphicx,tabularx,color}
\usepackage{amssymb,textcomp,mathtools, amsthm}
\usepackage{bm,upgreek,algorithm,hyperref}
\usepackage{multirow,booktabs,hhline,array}
\usepackage{cite,url,makecell,setspace}
\RequirePackage{doi}

\def\thline{\noalign{\hrule height 1.0pt}}

\renewcommand{\vec}[1]{\bm{\mathrm{#1}}}

\theoremstyle{definition}

\theoremstyle{plain}

\theoremstyle{remark}

\title{Group Communication with Context Codec for Lightweight Source Separation}

\author{Yi~Luo, Cong~Han, Nima~Mesgarani}

\begin{document}
\maketitle
\setlength{\abovedisplayskip}{2pt}
\setlength{\belowdisplayskip}{2pt}
\setlength{\abovedisplayshortskip}{2pt}
\setlength{\belowdisplayshortskip}{2pt}

\begin{abstract}
Despite the recent progress on neural network architectures for speech separation, the balance between the model size, model complexity and model performance is still an important and challenging problem for the deployment of such models to low-resource platforms. In this paper, we propose two simple modules, group communication and context codec, that can be easily applied to a wide range of architectures to jointly decrease the model size and complexity without sacrificing the performance. A group communication module splits a high-dimensional feature into groups of low-dimensional features and captures the inter-group dependency. A separation module with a significantly smaller model size can then be shared by all the groups. A context codec module, containing a context encoder and a context decoder, is designed as a learnable downsampling and upsampling module to decrease the length of a sequential feature processed by the separation module. The combination of the group communication and the context codec modules is referred to as the GC3 design. Experimental results show that applying GC3 on multiple network architectures for speech separation can achieve on-par or better performance with as small as 2.5\% model size and 17.6\% model complexity, respectively.
\end{abstract}


\section{Introduction}
\label{sec:intro}
Recent developments in the neural network architectures have significantly advanced the state-of-the-art for source separation performance. While researchers have shown that large neural networks can achieve superb performance on various machine learning problems, one important topic for network design, especially for the source separation problem, is to constrain the model size and complexity so that such models can be properly deployed to low-resource platforms such as mobile and hearable devices.

Tremendous efforts have been made to propose novel model architectures and model compression techniques. Early deep neural networks used for source separation contained stacked recurrent layers such as LSTM layers with a relatively large number of hidden units \cite{yu2017permutation, kolbaek2017multitalker, isik2016single, luo2017speaker}, and the corresponding model sizes were typically over tens of millions of parameters with high model complexity. Convolutional neural networks (CNNs) have also been explored in both time-frequency domain \cite{chandna2017monoaural, takahashi2017multi, yin2020phasen, takahashi2020d3net} and time domain \cite{stoller2018wave, venkataramani2018end, luo2019conv, lluis2019end, pandey2019new, zhang2020furcanext, tzinis2020sudo, zeghidour2020wavesplit}, and researchers have begun to focus on decreasing the model size and complexity while maintaining or improving the performance. Moreover, the combination of recurrent and convolutional layers has also been a popular topic for real-time model design, and various convolutional recurrent networks have been proposed \cite{sun2017multiple, naithani2017low, zhao2018convolutional, tan2018convolutional, hu2020dccrn}. Better layer organization within the network have also been investigated \cite{luo2020dual, nachmani2020voice, chen2020dual, kinoshita2020multi}, which can further decrease the overall model size and maintain the separation fidelity. Beyond directly designing smaller models, neural architecture search (NAS) techniques have also been utilized to automatically search for compact architectures for speech-related tasks \cite{mazzawi2019improving, hu2020neural}, teacher-student learning methods have been explored for obtaining low-latency separation models \cite{aihara2019teacher}, quantization and binarization algorithms have been studied for low-resource separation systems \cite{kim2018bitwise, kim2019incremental, kim2020boosted}, and network pruning and distillation strategies can further be applied to decrease the model size \cite{hinton2015distilling, luo2017thinet}. However, compared with directly designing lightweight architectures, existing model compression or quantization techniques typically introduce different levels of degradation on the model performance \cite{simons2019review}, and the tradeoff between the complexity and performance drop needs to be carefully considered. Thus, finding smaller and better architectures is still preferred.

Our recent study introduced \textit{group communication (GroupComm)} \cite{luo2020ultra}, a module motivated by subband and multiband processing network architectures such as frequency-LSTM (F-LSTM) \cite{li2015lstm, wang2017joint, zhang2017deep, li2019multichannel}, which can easily change a separation model into a lightweight counterpart. GroupComm splits a high-dimensional feature, such as a spectrum, into groups of low-dimensional features, such as subband spectra, and uses the same separation model across all the groups for weight sharing. Another inter-group module is applied to capture the dependencies within the groups, so that the processing of each group always depends on the global information available. Compared with conventional F-LSTM or other similar architectures that explicitly model time and frequency dependencies where the subband features are concatenated back to the fullband feature \cite{li2016exploring, sainath2016modeling, xu2018single}, GroupComm does not perform such concatenation but simply applies a small module to communicate across the groups. Moreover, the low-dimensional features enable the use of a smaller module, e.g., a CNN or RNN layer, than the original high-dimensional feature, and together with weight sharing the total model size can be significantly reduced. Experimental results showed that the GroupComm-equipped model can achieve on-par performance with the baseline model in the noisy reverberant speech separation task with a 2.8\% model size and 43.5\% multiply-accumulate (MAC) operations \cite{whitehead2011precision}, a common metric for evaluating model complexity.

It is worth noting that although GroupComm can drastically decrease the model size, model complexity is still relatively high. Moreover, memory footprints in several GroupComm-equipped models with small group sizes are higher than those in the baselines due to the additional computation introduced by the GroupComm modules, which may not only pose constraints on its applications but also increase the training cost. In this paper, we introduce a \textit{context codec} module to help GroupComm maintain the performance while further decreasing the number of MAC operations, accelerating the training speed and alleviating the memory consumption in both training and inference time. A context codec module contains a context encoder and a context decoder, where the context encoder summarizes the temporal context of local features into a single feature representing the global characteristics of the context, and a context decoding module transforms the compressed feature back to the context features. Squeezing the input contexts into higher-level representations corresponds to a nonlinear downsampling step that generates context-level embeddings and significantly decreases the length of a feature sequence. Note that compared with other architectures that perform iterative downsampling and upsampling steps \cite{stoller2018wave, tzinis2020sudo}, the context codec is only applied once and all remaining modeling steps are applied on the downsampled features, which enables a smaller memory footprint and faster training speed. We call the combination of \textbf{\textit{G}}roup\textbf{\textit{C}}omm and \textbf{\textit{C}}ontext \textbf{\textit{C}}odec the \textbf{\textit{GC3}} design. 

The original GroupComm module applied a bidirectional LSTM (BLSTM) layer for inter-group processing. In this paper, we also explore different architectures for the GroupComm module and investigate the effect of different hyperparameters in the system configuration. Moreover, to validate the effect of GC3 on different network architectures, we select three other separation modules beyond the original dual-path RNN (DPRNN) baseline \cite{luo2020dual} applied in our previous work on GroupComm: two other CNN-based architectures, namely the temporal convolutional network (TCN) \cite{luo2019conv} and the sudo rm -rf network \cite{tzinis2020sudo}, and one transformer-based architecture, namely the dual-path transformer network (DPTNet) \cite{chen2020dual}. Experimental results show that the GC3-equipped DPRNN can achieve on-par performance with the baseline DPRNN with 4.7\% model size and 17.6\% MAC operations, the GC3-equipped CNN-based models can significantly improve the overall performance with as few as 2.5\% model size and 33.7\% MAC operations, and the GC3-equipped transformer-based model can maintain on-par performance with 4.6\% model size and 17.7\% MAC operations. These results indicate that GC3 has the potential to be a plug-and-play module suitable for a wide range of architectures.

The rest of the paper is organized as follows. Section~\ref{sec:GC3} introduces the GC3 design and discusses the applicable model architectures. Section~\ref{sec:config} provides the experiment configurations. Section~\ref{sec:result} analyzes the experimental results. Section~\ref{sec:conclusion} concludes the paper.

\section{Group Communication with Context Codec}
\label{sec:GC3}
\begin{figure*}[!t]
	\small
	\centering
	\includegraphics[width=1.8\columnwidth]{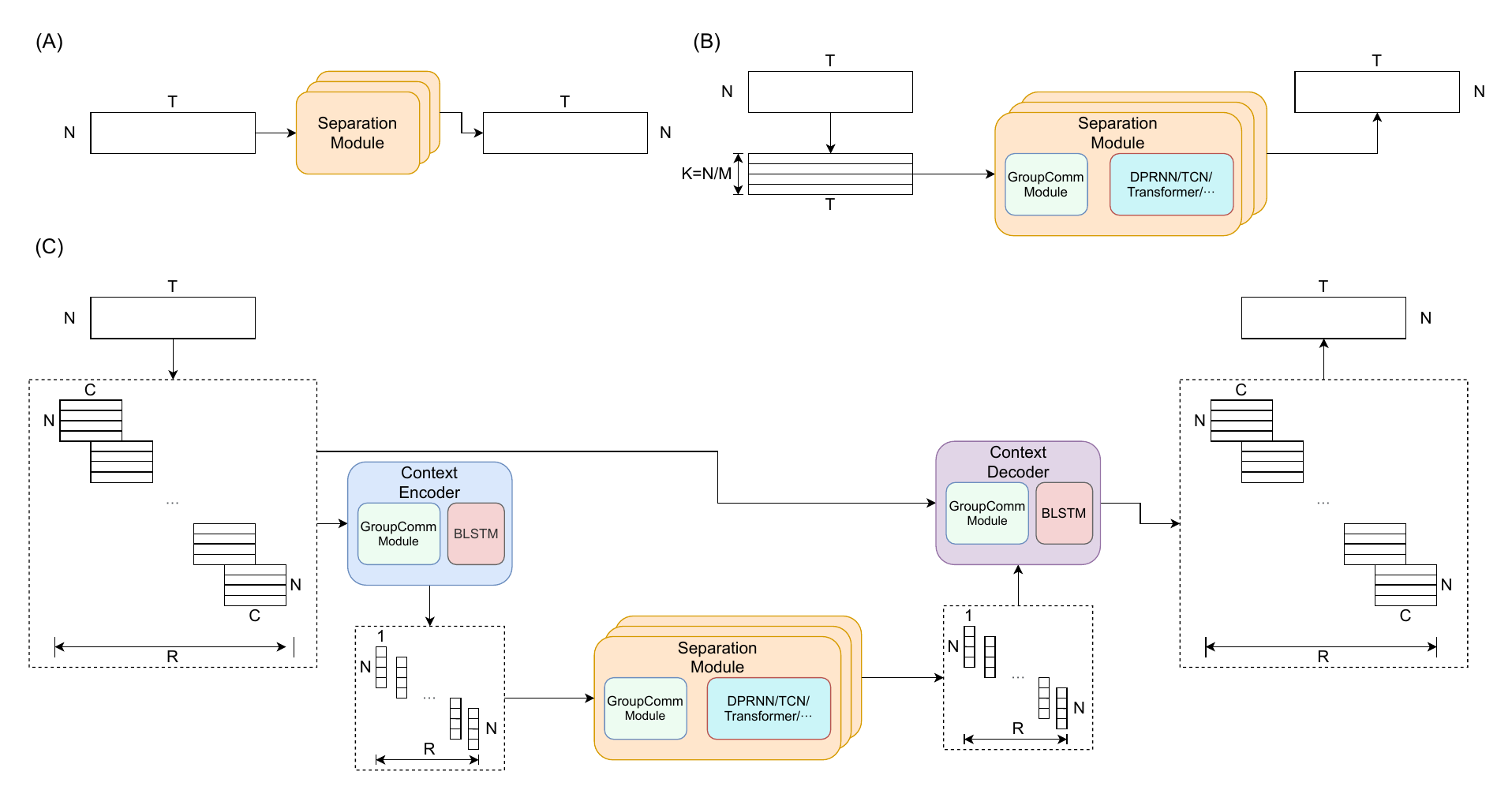}
	\caption{Flowcharts for (A) standard sequence processing pipeline with a large sequence modeling module; (B) GroupComm-equipped pipeline, where the features are split into groups with a GroupComm module for inter-group communication. A smaller module for sequence modeling is then shared by all groups; (C) GC3-equipped pipeline, where the sequence is first segmented into local context frames, and each context is encoded into a single feature. The sequence of summarized features is passed to a GroupComm-equipped module in (B). The transformed summarized features and the original local context frames are passed to a context decoding module and an overlap-add operation to generate the output with the same size as the input sequence.}
	\label{fig:flowchart}
\end{figure*}

\subsection{Standard Pipeline for Speech Separation}
\label{sec:standard-pipeline}

We start with a brief overview of a standard pipeline for speech separation. Given the input mixture waveforms $\vec{x} \in \mathbb{R}^{P\times L}$ where $P$ denotes the number of available channels and $L$ denotes the waveform length, an encoding transformation is typically applied to each channel to generate a 2-D representation:
\begin{align}
    \vec{H}_i = E(\vec{x}_i), \quad i = 1, \ldots, P
\end{align}
where $\vec{H}_i \in \mathbb{C}^{N\times T}$ is the 2-D representation at channel $i$, where $N$ denotes the representation dimension and $T$ denotes the number of frames or time steps, and $E(\cdot)$ is the encoding transform. We assume a single-channel separation setting with $P=1$ and drop the subscript where there is no ambiguity. For frequency-domain methods, $E(\cdot)$ denotes the short-time Fourier transform (STFT). For certain time-domain methods such as the time-domain audio separation network (TasNet) \cite{luo2018tasnet, luo2019conv}, $E(\cdot)$ denotes a 1-D convolutional layer and $\vec{H}_i$ becomes a real-valued representation. 

$\vec{H}$ is then passed to a separation module to generate $X$ outputs $\{\hat{\vec{H}}_c\}_{c=1}^X \in \mathbb{C}^{N\times T}$ corresponding to the $X$ target representations:
\begin{align}
    \{\hat{\vec{H}}_c\}_{c=1}^X = T(\vec{H})
\end{align}
where $T(\cdot)$ is the mapping defined by the separation module. In most recent speech separation networks, $T(\cdot)$ is defined by certain sequence modeling neural networks such as deep long-short term memory (LSTM) networks \cite{hochreiter1997long}, 1-D convolutional neural networks \cite{luo2019conv}, or Transformers \cite{chen2020dual}.

A decoding transform $D(\cdot)$ is then applied to all $\hat{\vec{H}}_c$ to reconstruct the estimated target waveforms $\{\vec{s}_c\}_{c=1}^X \in \mathbb{R}^{1\times L}$:
\begin{align}
    \vec{s}_c = D(\hat{\vec{H}}_c), \quad c = 1, \ldots, X
\end{align}
Similarly, $D(\cdot)$ denotes the inverse short-time Fourier transform (ISTFT) in frequency-domain methods and denotes a 1-D transposed convolutional layer in TasNet-style methods. Figure~\ref{fig:flowchart} (A) presents the separation part in this pipeline.

\subsection{Group Communication}
\label{sec:GroupComm-design}

Every frame in $\vec{H}$, denoted by $\vec{h} \in \mathbb{R}^N$, can be decomposed into $K$ groups of lower-dimensional feature vectors $\{\vec{g}^i\}_{i=1}^K$ with $\vec{g}^i \in \mathbb{R}^M$. When there is no overlap between the groups, we have $N=KM$. A \textbf{group communication (GroupComm)} module is applied across the group of vectors to capture the inter-group dependencies:
\begin{align}
	\{\hat{\vec{g}}^i\}_{i=1}^K = F(\{\vec{g}^i\}_{i=1}^K)
\label{eqn:groupcomm}
\end{align}
where $\hat{\vec{g}}^i \in \mathbb{R}^P$ is the transformed feature vector for group $i$, and $F(\cdot)$ is the mapping function defined by the module. Instead of concatenating $\{\hat{\vec{g}}^i\}_{i=1}^K$ back to an $N$-dimensional feature, $\hat{\vec{g}}^i$ are independently passed to a smaller separation module to save the model size and complexity. In other words, all groups share a same separation module whose width can be significantly smaller than the original separation module applied to $\vec{H}$. Figure~\ref{fig:flowchart} (B) presents the flowchart for a GroupComm-equipped separation module.

\subsection{GroupComm with Context Codec}
\label{sec:GC3-design}

In TasNet-style time-domain methods, the 1-D convolutional waveform encoder contains a much smaller window length than standard frequency-domain methods, and it has been studied that shorter window length can lead to a better separation performance \cite{luo2019conv, luo2020dual}. However, a shorter window length leads to a larger $T$ which increases both the computational complexity and the modeling difficulty of the separation module. Our prior study showed that although GroupComm allowed a smaller separation module, the computation introduced by the GroupComm modules can still be large since the GroupComm module has to be applied to all the frames in $\vec{H}$ \cite{luo2020ultra}. How to decrease $T$ while maintaining the model performance is thus an important question.

A \textbf{context codec} is proposed here as a pair of encoding and decoding modules, which compress the context of feature vectors into a single summarization vector and decompress the vector back to a context, respectively. A context encoder splits $\vec{H}$ along the temporal dimension into blocks $\{\vec{D}^i\}_{i=1}^R \in \mathbb{R}^{N\times C}$, where $C$ denotes the context size and $R$ denotes the number of context blocks. Each $\vec{D}^i$ is then encoded into a single vector $\vec{p}^i \in \mathbb{R}^{W}$ by the context encoder, resulting in a sequence of vectors $\vec{P} \triangleq \{\vec{p}^i\}_{i=1}^R \in \mathbb{R}^{W\times R}$ with $R\ll T$. Any separation module can then be applied to $\vec{P}$ instead of $\vec{H}$ to save the computational cost. The transformed sequence of features are denoted as $\hat{\vec{P}} \in \mathbb{R}^{W\times R}$, and a context decoding module adds $\hat{\vec{p}}^i$ to each time step in $\vec{D}^i$ and applies a nonlinear transformation to generate $\hat{\vec{D}}^i \in \mathbb{R}^{N\times C}$ for context $i$. Overlap-add is then applied on $\{\hat{\vec{D}}^i\}_{i=1}^R$ to form the sequence of features $\hat{\vec{H}} \in \mathbb{R}^{N\times T}$ of the original length. 

We select a deep residual BLSTM network for both the context encoder and decoder in our configuration. The deep residual BLSTM networks contained stacked BLSTM layers, where each BLSTM layer contains a linear projected layer connected to its output to match the input and output feature dimensions. A layer normalization (LayerNorm) operator \cite{ba2016layer} is added to the transformed output, a residual connection is added between the input to the BLSTM layer and the LayerNorm-normalized output, and the feature is then served as the input for the next layer. In the context encoder, a context block $\vec{D}^i$ is passed to the GroupComm-equipped deep residual BLSTM network to generate a transformed sequence of features $\vec{Q}^i \in \mathbb{R}^{N\times C}$, and a mean-pooling operation is applied on $\vec{Q}^i$ across the temporal dimension to obtain $\vec{p}^i$. In the context decoder, $\hat{\vec{p}}^i$ is added to each time step in $\vec{D}^i$ and passed to the GroupComm-equipped deep residual BLSTM network to generate the final output $\hat{\vec{D}}^i$. To save the computational cost in the context codec, GroupComm is also applied to the deep residual BLSTM networks in the same way as Eq.~\ref{eqn:groupcomm}. This combination of GroupComm and context codec gives us the \textbf{GC3} design, and Figure~\ref{fig:flowchart} (C) provides the flowchart for the GC3-equipped separation pipeline.

Note that there is no guarantee that the context encoding and decoding modules are reconstructing the original input features as a ``codec'' typically does, and here, we borrow the name of codec simply to represent the encoding and decoding properties of the two modules.

\subsection{Discussions}

Splitting a high-dimensional feature into low-dimensional sub-features has also been investigated in architectures for computer vision tasks \cite{gao2019res2net, zhang2020resnest, han2020ghostnet}. GroupComm shares the same principle as those designs for exploring the nonlinear dependendies at the sub-feature level, while GroupComm removes the feature concatenation operation and assumes that a small module shared by all the groups is adequate to preserve the model capacity given the inter-group modeling step.

As discussed in \cite{luo2020ultra}, the network width in the separation module in Figure~\ref{fig:flowchart} depends on the group size $K$. Empirically, we linearly decrease the network width as $K$ increases. To provide an example, we assume that for a standard separation module in Figure~\ref{fig:flowchart} (A) with $N$-dimensional input features, the network width of the separation module is denoted by $Q$ and both $N$ and $Q$ are multiples of $K$. By assuming that there is no overlap between groups, the input dimension for both the GroupComm module and the separation submodule (denoted by possible choices of architectures in Figure~\ref{fig:flowchart} (B) and (C)) is defined by $M=N/K$, and we can empirically set the network width to $Q/K$. \cite{luo2020ultra} already conducted experiments on the effect of hyperparameters on the overall model performance and MAC operations, and it was observed that to achieve on-par performance with the baseline, a smaller $K$ requires a deeper separation module. We show in Section~\ref{sec:result} that the GC3-equipped pipeline requires a slightly deeper separation module due to the existence of the context codec modules; however, the total number of MAC operations is fewer due to the shorter sequence length $R$ after context encoding.

Context information is widely used as auxiliary information to assist the modeling of a center frame in a sequential input. While many existing studies use a plain concatenation of context features \cite{xu2014regression, araki2015exploring, yu2017permutation}, context codec modules have also been investigated in various studies to learn a compact representation \cite{chen2015long, pathak2016context, mehri2016samplernn, marafioti2019context}. Specifically, \cite{luo2020implicit} applied a context codec for the multichannel separation task to learn a set of context-aware filters; however, the context codec did not decrease the sequence length $T$ but was only used to capture context-aware information. The computational cost in \cite{luo2020implicit} is thus even higher than that without context codec. The main role of the context codec in the GC3 framework is to decrease the computational cost in the actual sequence modeling module by decreasing the sequence length.

\section{Experiment configurations}
\label{sec:config}
\subsection{Data Simulation}

We use the same dataset proposed in \cite{luo2020end} for the single-channel noisy reverberant speech separation task. The simulated dataset contains 20000, 5000 and 3000 4-second long utterances sampled at 16 kHz sample rate for training, validation and test sets, respectively. For each utterance, two speech signals and one noise signal are randomly selected from the 100-hour Librispeech subset \cite{panayotov2015librispeech} and the 100 Nonspeech Corpus \cite{web100nonspeech}, respectively. The overlap ratio between the two speakers is uniformly sampled between 0\% and 100\%, and the two speech signals are shifted accordingly and rescaled to a random relative SNR between 0 and 5 dB. The relative SNR between the power of the sum of the two clean speech signals and the noise is randomly sampled between 10 and 20 dB. The transformed signals are then convolved with the room impulse responses simulated by the image method \cite{allen1979image} using the gpuRIR toolbox \cite{diaz2020gpurir} for all microphones. The length and width of all the rooms are randomly sampled between 3 and 10 meters, and the height is randomly sampled between 2.5 and 4 meters. The reverberation time (T60) is randomly sampled between 0.1 and 0.5 seconds. After convolution, the echoic signals are summed to create the mixture for each microphone. The data simulation scripts are publicly available online\footnote{\url{https://github.com/yluo42/TAC}}.

\subsection{Model Configurations}

\subsubsection{Separation Pipeline Configurations}

We test GC3 on the single-channel separation task, where only the first channel in the aforementioned dataset is used. We apply the TasNet-style time-domain separation configuration where a linear learnable 1-D convolutional encoder is used to learn the 2-D sequential representations $\vec{H}$, and a linear learnable 1-D transposed convolutional decoder is used to reconstruct the waveforms. The separation is done by estimating a set of multiplicative masks applied to the 2-D representation of the input mixture. We use a 2~ms window size (32 samples at 16 kHz sample rate) in the 1-D convolutional encoder and decoder for all experiments, and use ReLU nonlinearity as the activation function for the mask estimation layer. In models with GroupComm, the mask estimation layer is applied to different groups in parallel.

\subsubsection{GroupComm Module Configurations}
\label{sec:GroupComm-network}

The original GroupComm module proposed in \cite{luo2020ultra} applied a simple bidirectional LSTM (BLSTM) layer. However, the assumption that a sequential order exists in the features in different groups is ad-hoc. Here we investigate two other architectures for the GroupComm module:
\begin{enumerate}
\item \textit{Transform-average-concatenate (TAC)} \cite{luo2020end}: TAC was proposed for the multichannel speech separation task with ad-hoc microphone arrays where no microphone indexing or geometry information is known in advance. The design particularly matches our need in the GroupComm module where ``group indices'', i.e., the sequential order of the features in different groups, does not exist. For the group of features $\{\vec{g}^i\}_{i=1}^K$, a fully-connected (FC) layer with parametric rectified linear unit (PReLU) activation \cite{he2015delving} is applied for the transformation step:
\begin{align}
    \vec{f}^i = P(\vec{g}^i)
\end{align}
where $P(\cdot)$ is the mapping function defined by the first FC layer and $\vec{f}^i \in \mathbb{R}^{D}$ denotes the output for group $i$. All $\vec{f}^i$ are then averaged and passed to the second FC layer with PReLU activation for the averaging step:
\begin{align}
    \hat{\vec{f}} = R(\frac{1}{K}\sum_{i=1}^{K} \vec{f}^i)
\end{align}
where $R(\cdot)$ is the mapping function defined by the second FC layer and $\hat{\vec{f}} \in \mathbb{R}^{D}$ is the output for this step. $\hat{\vec{f}}$ is finally concatenated with the output of the transformation step, $\vec{f}^i$, and passed to a third FC layer with PReLU activation to generate the final output $\hat{\vec{g}}^i$:
\begin{align}
    \hat{\vec{g}}^i = S([\vec{f}^i; \hat{\vec{f}}])
\end{align}
where $S(\cdot)$ is the mapping function defined by the third FC layer and $[\vec{x};\vec{y}]$ denotes the concatenation operation of vectors $\vec{x}$ and $\vec{y}$. A residual connection is finally added between the module input $\vec{g}^i$ and output $\hat{\vec{g}}^i$.

\item \textit{Multi-head Self Attention (MHSA)} \cite{vaswani2017attention}: MHSA is widely used in various sequence modeling tasks and has already proven its effectiveness in multiple speech-related problems \cite{vaswani2017attention, dong2018speech, li2019neural, tsunoo2019transformer, karita2019comparative}. MHSA explicitly models the relationship between each pair of group features and thus can capture sequence-level dependencies. Following the standard definition of MHSA, we rewrite the concatenation of group features $\{\vec{g}^i\}_{i=1}^K$ as a matrix $\vec{G} \in \mathbb{R}^{K\times M}$ and apply a MHSA layer:
\begin{align}
    \text{H}_n &= \text{Softmax}(\frac{\vec{Q}_n\vec{K}_n^T}{\sqrt{d_k}})\vec{V}_n \\
    \vec{G}^\prime_n &= [\text{H}_1;\cdots;\text{H}_n]\vec{W}^o
\end{align}
where $n$ is the number of attention heads, $\vec{W}_n^q,\vec{W}_n^k,\vec{W}_n^v \in \mathbb{R}^{M\times d_k}$ are the linear transformation matrices for head $n$, $\vec{Q}_n=\vec{G}\vec{W}_n^q$, $\vec{K}_n=\vec{G}\vec{W}_n^k$, and $\vec{V}_n=\vec{G}\vec{W}_n^v$ are the linear transformations for query, key and value, respectively, $\text{H}_n \in \mathbb{R}^{K\times d_k}$ is the output at head $n$, and $\vec{W}^o$ is the linear transformation matrix for the output. Both the concatenation operation and the Softmax nonlinearity are applied across the attention heads. $\vec{G}^\prime$ is then passed to an FC layer with PReLU activation for further transformation, and another FC layer follows to generate the final output with the same shape as input $\vec{G}^{\prime\prime} \in \mathbb{R}^{K\times M}$.
\end{enumerate} 

We select the hyperparameters for the three architectures such that they all have a same number of parameters. The configuration of the modules is introduced in Section~\ref{sec:GroupComm-model}.

\subsubsection{Separation Module Configurations}
\label{sec:separation-network}

To investigate the effect of GC3 on different network architectures, we adopt 4 types of separation modules in our experiments:

\begin{enumerate}
    \item \textit{Dual-path RNN (DPRNN)} \cite{luo2020dual}: A dual-path RNN (DPRNN) segments a sequence of features into overlapping blocks and iteratively applies an intra-block BLSTM layer and an inter-block BLSTM layer to capture the local and global dependencies, respectively. DPRNN is also the architecture selected in our prior work on GroupComm \cite{luo2020ultra}.
    
    \item \textit{Temporal convolutional network (TCN)} \cite{luo2019conv}: A TCN block contained a depthwise-separable convolution layer \cite{chollet2016xception} with PReLU nonlinearity and a LayerNorm operation, where the convolution layer contains an exponentially increased dilation factor to increase the overall receptive field of the TCN.
    
    \item \textit{Sudo rm -rf} \cite{tzinis2020sudo}: Sudo rm -rf proposed a U-net style convolutional block where multiple levels of downsampling and upsampling layers were applied to extract features at different scales. Each downsampling layer contained a depthwise separation convolution operation similar to Conv-TasNet, and each upsampling layer contained a bilinear interpolation operation.
    
    \item \textit{Dual-path Transformer (DPTNet)} \cite{chen2020dual}: DPTNet replaced the BLSTM layers in DPRNN with modified Transformer layers \cite{vaswani2017attention}, where the fully connected layer in the default transformer encoder layer \cite{vaswani2017attention} was replaced by an LSTM layer to learn the positional information in the sequence.
\end{enumerate}

We encourage the readers to refer to the corresponding literature for details about these architectures. All separation modules share a same configuration for the encoder and decoder and the mask estimation layer. The implementation of all models with GC3 is available online\footnote{\url{https://github.com/yluo42/GC3}}.

\subsection{Training Configurations}

All models are trained for 100 epochs with the Adam optimizer with an initial learning rate of 0.001. Negative signal-to-noise ratio (SNR) is used as the training objective for all models. The learning rate is decayed by 0.98 for every two epochs. Gradient clipping by a maximum gradient norm of 5 is always applied for proper convergence. Early stopping is applied when no best validation model is found for 10 consecutive epochs. No other training tricks or regularization techniques are used. Auxiliary autoencoding training (A2T) is applied to enhance the robustness on this reverberant environment configuration \cite{luo2020distortion}, where the multiplicative mask estimated for each target is also applied to the corresponding direct-path signal, and the auxiliary autoencoding loss is applied to ensure that the multiplicative mask maintains the direct-path signal. For more details about A2T, we recommend the readers to refer to the original literature.

\subsection{Evaluation metrics}

In the experiments for hyperparameter search, we report the SI-SDR score \cite{le2019sdr} for the evaluation of the separation performance, and the model size and the number of MAC operations (MACs) as metrics for model complexity. MACs for all models are calculated by an open-source toolbox\footnote{\url{https://github.com/Lyken17/pytorch-OpCounter}}. In the experiments for comparing the performance of different separation modules with and without GC3, we also report the wideband perceptual evaluation of speech quality (PESQ) score \cite{rix2001perceptual} and the short-time objective intelligibility measure (STOI) \cite{taal2010short}.

\section{Results and analysis}
\label{sec:result}
\begin{table}[!t]
	\scriptsize
	\centering
	\caption{Hyperparameters and their notations in DPRNN-based models.}
	\begin{tabular}{c|c}
		\thline
		Hyperparameter & Notation \\
		\thline
		Number of groups & $K$ \\
		Group size & $M$ \\
		Number of encoder filters & $N$ \\
		Input / hidden dimensions in all BLSTM layers & $H_i$ / $H_o$ \\
		Number of DPRNN blocks & $L_s$ \\
		Number of context codec layers & $L_c$ \\
		Context size (in frames) & $C$ \\
		DPRNN block size (in frames) & $B$ \\ 
		\thline
	\end{tabular}
	\label{tab:param}
\end{table}

\begin{table*}[!t]
	\scriptsize
	\centering
	\caption{Comparison of DPRNN, GroupComm-DPRNN and GC3-DPRNN TasNet models with different hyperparameter configurations. MACs are calculated on 4-second mixtures.}
	\begin{tabular}{c|c|c|c|c|c|c|c|c|c|c|c}
		\thline
		Model & $K$ & $M$ & $N$ & $H_{i}$ / $H_{o}$ & $L_s$ & $L_c$ & $C$ & $B$ & SI-SDR (dB) & Model size & MACs \\
		\thline
		DPRNN & 1 & 128 & 128 & 64 / 128 & 6 & -- & -- & 100 & 9.0 & 2.6M & 22.1G \\
		\thline
		\multirow{11}{*}{GroupComm-DPRNN} & 2 & 64 & 128 & 64 / 128 & 4 & \multirow{11}{*}{--} & \multirow{11}{*}{--} & \multirow{11}{*}{100} & 9.5 & 2.6M (99.4\%) & 43.4G (196.4\%) \\
		\cline{2-6}\cline{10-12}
		& 4 & 32 & 128 & 32 / 64 & 4 & & & & 9.4 & 663.0K (25.3\%) & 22.4G (101.4\%) \\
		\cline{2-6}\cline{10-12}
		& 8 & 16 & 128 & 16 / 32 & 4 & & & & 8.9 & 175.5K (6.7\%) & 11.9G (53.8\%) \\
		\cline{2-6}\cline{10-12}
		& \multirow{4}{*}{16} & \multirow{2}{*}{8} & \multirow{2}{*}{128} & \multirow{2}{*}{8 / 16} & 4 & & & & 8.1 & 51.9K (2.0\%) & 6.6G (29.9\%) \\
		& & & & & 6 & & & & \textbf{8.9} & \textbf{73.5K (2.8\%)} & \textbf{9.6G (43.4\%)} \\
		\cline{3-6}\cline{10-12}
		& & \multirow{2}{*}{16} & \multirow{2}{*}{256} & \multirow{2}{*}{16 / 32} & 2 & & & & 8.1 & 100.7K (3.8\%) & 12.4G (56.1\%) \\
		& & & & & 4 & & & & 9.7 & 183.9K (7.0\%) & 23.7G (107.2\%) \\
		\cline{2-6}\cline{10-12}
		& \multirow{4}{*}{32} & \multirow{2}{*}{4} & \multirow{2}{*}{128} & \multirow{2}{*}{4 / 8} & 6 & & & & 7.6 & 26.0K (1.0\%) & 5.7G (25.8\%) \\
		& & & & & 10 & & & & \textbf{8.5} & \textbf{37.6K (1.4\%)} & \textbf{9.1G (41.2\%)} \\
		\cline{3-6}\cline{10-12}
		& & \multirow{2}{*}{8} & \multirow{2}{*}{256} & \multirow{2}{*}{8 / 16} & 2 & & & & 7.9 & 38.7K (1.5\%) & 7.2G (32.6\%) \\
		& & & & & 4 & & & & 8.6 & 60.3K (2.3\%) & 13.2G (59.7\%) \\
		\thline
		\multirow{10}{*}{GC3-DPRNN} & \multirow{4}{*}{4} & \multirow{4}{*}{32} & \multirow{4}{*}{128} & \multirow{4}{*}{32 / 64} & \multirow{4}{*}{4} & \multirow{4}{*}{1} & 32 & 24 & 8.9 & \multirow{4}{*}{881.5K (33.7\%)} & 9.2G (41.6\%) \\
		& & & & & & & 16 & 32 & 8.8 & & 10.6G (48.0\%) \\
		& & & & & & & 8 & 50 & 8.8 & & 13.3G (60.2\%) \\
		& & & & & & & 4 & 64 & 9.3 & & 18.6G (84.2\%) \\
		\cline{2-12}
		& \multirow{2}{*}{8} & \multirow{2}{*}{16} & \multirow{2}{*}{128} & \multirow{2}{*}{16 / 32} & \multirow{2}{*}{6} & 1 & \multirow{6}{*}{32} & \multirow{6}{*}{24} & 8.7 & 314.4K (12.0\%) & 5.4G (24.4\%) \\
		& & & & & & 2 & & & 9.3 & 369.9K (14.1\%) & 8.1G (36.7\%) \\
		\cline{2-7}\cline{10-12}
		& \multirow{2}{*}{16} & 8 & 128 & 8 / 16 & 8 & 2 & & & \textbf{8.9} & \textbf{124.1K (4.7\%)} & \textbf{5.4G (24.4\%)} \\
		& & 16 & 256 & 16 / 32 & 6 & 1 & & & 9.0 & 322.9K (12.3\%) & 10.9G (49.3\%) \\
		\cline{2-7}\cline{10-12}
		& \multirow{2}{*}{32} & 4 & 128 & 4 / 8 & 14 & \multirow{2}{*}{2} & & & \textbf{8.3} & \textbf{57.1K (2.2\%)} & \textbf{3.6G (16.3\%)} \\
		& & 8 & 256 & 8 / 16 & 8 & & & & 9.2 & 132.6K (5.1\%) & 10.7G (48.4\%) \\
		\thline
	\end{tabular}
	\label{tab:result}
\end{table*}

\begin{table}[!t]
	\scriptsize
	\centering
	\caption{Comparison of GC3-DPRNN models with different model architectures for GroupComm.}
	\begin{tabular}{c|c|c|c|c}
		\thline
		GroupComm module & $K$ & SI-SDR (dB) & Model size & MACs \\
		\thline
		\multirow{2}{*}{BLSTM} & 16 & 8.9 & 124.1K & 5.4G (24.4\%) \\
		& 32 & 8.3 & 57.1K & 3.6G (16.3\%) \\
		\hline
		\multirow{2}{*}{TAC} & 16 & \textbf{9.1} & \textbf{123.8K} & \textbf{3.9G (17.6\%)} \\
		& 32 & \textbf{8.6} & \textbf{56.3K} & \textbf{2.6G (11.8\%)} \\
		\hline
		\multirow{2}{*}{MHSA} & 16 & 8.9 & 123.7K & 4.5G (20.3\%) \\
		& 32 & 8.2 & 56.5K & 3.0G (13.4\%) \\
		\thline
	\end{tabular}
	\label{tab:GroupComm}
\end{table}

In this section, we first evaluate the performance of GroupComm-equipped DPRNN-TasNet and GC3-equipped DPRNN-TasNet by comparing to the baseline DPRNN-TasNet. We then select the best GC3 configuration and investigate the effect of different configurations in the GroupComm module. After that, we change DPRNN to three other types of separation modules and validate the performance of GC3 on various architectures.

\subsection{Experimental Results on GC3-DPRNN}

Our previous study \cite{luo2020ultra} conducted experiments on the effect of GroupComm in the DPRNN-TasNet architecture. A residual BLSTM layer identical to the one used in the context codec was selected for the GroupComm module in \cite{luo2020ultra}, and we present the experimental results on the comparison between original DPRNN-TasNet, the GroupComm-equipped DPRNN-TasNet, and the GC3-equipped DPRNN-TasNet. Table~\ref{tab:param} shows the hyperparameter notations used in the DPRNN-based models. Note that $H_i$ and $H_o$ represent the input and output dimensions of all the BLSTM layers throughout the network since all context codec, GroupComm and DPRNN blocks use residual BLSTM as basic building units. Table~\ref{tab:result} lists the separation performance and the corresponding model size and complexity for the three types of models. The first half of the table is identical to the experiment results presented in \cite{luo2020ultra}. For the standard DPRNN model, a linear bottleneck FC layer is applied to $\vec{H}$ to transform the $N$-dimensional feature to $H_i$-dimensional feature. For all GroupComm-equipped models, $\vec{H}$ is split into groups and then directly sent to the context encoder without the bottleneck layer.

We first notice that when the number of groups $K$ is small (e.g. $K\leq4$), the GroupComm-DPRNN models can achieve higher performance than plain DPRNN with smaller model size at the cost of an on-par or higher model complexity. This is due to the extra MAC operations introduced by the GroupComm module. A larger number of groups $K$ leads to fewer MAC operations; however, the depth of the model has to be modified accordingly to maintain the performance. We explore different hyperparameter settings such that for each value of $K$ we obtain a model with less than 5\% relative performance degradation. Among all the GroupComm-DPRNN models, we find that the model with $K=16$, $N=128$ and $L_s=6$ achieves on-par performance as the standard DPRNN model with 2.8\% model size and 43.4\% MAC operations, and the model with $K=32$, $N=128$ and $L_s=10$ only has 5\% performance degradation with 1.4\% model size and 41.2\% MAC operations. These models show that GroupComm is effective in decreasing both the model size and complexity without sacrificing the performance. Moreover, we also conduct experiments on the effect of model width in terms of number of encoder filters $N$ and network width $Q$ (i.e., $H_o$ in DPRNN-based models) and observe that increasing model width can improve the performance with a much shallower architecture; however, the model complexity can be relatively high. For example, a performance improvement of 0.7 dB can be achieved by $K=16$, $N=256$ and $L_s=4$, while its number of MAC operations is even higher than the baseline DPRNN. This indicates that when the computational cost is not a bottleneck, GroupComm can also be applied to improve the overall performance.

For the GC3-based models, we first investigate the balance between model complexity and performance with different context sizes $C$. A larger context size $C$ leads to fewer frames for the sequence modeling module, and a smaller DPRNN block size $B$ can be applied to save model complexity. We observe that models with $C=32, 16$ and $8$ have almost the same performance while differing greatly in complexity; hence, we choose $C=32$ for all other experiments. For $K=8$, we see that increasing the number of context codec layers can improve the separation performance, implying that a strong context codec is important. The on-par performance can be achieved by $K=16$, $N=128$, $L_s=8$ and $L_c=2$ with a 4.7\% model size and 24.4\% MAC operations, which saves 19\% more MAC operations compared with the GroupComm-only model. The GC3-based model with 5\% performance degradation has the configuration of $K=32$, $N=128$, $L_s=14$ and $L_c=2$, which saves 25\% more MAC operations than the GroupComm-only model. Such results prove that GC3-based models are more effective than GroupComm-only models thanks to the context compression operation. Similarly, increasing the model width can also lead to better overall performance with a shallower architecture at the cost of complexity, and in such configurations it is empirically better to keep the depth of the context codec according to the results for $K=16$.

\begin{table}[!t]
	\scriptsize
	\centering
	\caption{Effect of group overlap ratio on model complexity and separation performance in GC3-DPRNN models.}
	\begin{tabular}{c|c|c|c}
		\thline
		Group overlap & SI-SDR (dB) & Model size & MACs \\
		\thline
		0\% & 9.1 & \multirow{3}{*}{123.8K} & 3.9G (17.6\%) \\
		25\% & 9.0  & & 4.9G (22.0\%)  \\
		50\% & 9.4 & & 6.9G (31.3\%) \\
		\thline
	\end{tabular}
	\label{tab:overlap}
\end{table}

\subsection{Effect of Model Architectures for GroupComm}
\label{sec:GroupComm-model}

We then evaluate the effect of model architectures for GroupComm. We compare BLSTM with the two other models introduced in Section~\ref{sec:GroupComm-network}. For the TAC architecture, the hidden dimension $D$ is set to $3H_o$. For the MHSA architecture, we use 4 attention heads with the hidden dimension $d_k$ set to $M$. Such configurations are applied to match the overall model sizes with a residual BLSTM layer. We use the hyperparameter configurations of the two GC3-based models marked in bold in Table~\ref{tab:result} with $K=16$ and $32$, respectively. Table~\ref{tab:GroupComm} shows the comparison of the three architectures. We find that although MHSA achieves on-par performance as BLSTM in both configurations, TAC obtains even better performance with the fewest MAC operations. Since the number of MAC operations in TAC is fewer than those in both BLSTM and MHSA and the transformation and concatenation steps in TAC can be run in parallel across groups, we use TAC as the default module for GroupComm in all remaining experiments.

\subsection{Effect of Overlap between Groups}

The default group segmentation configuration in all experiments above assumes no overlap between groups. However, since a 50\% overlap is applied in both context codec and the sequence segmentation process in DPRNN, it is thus interesting to see whether adding overlap between groups can improve the performance. Table~\ref{tab:overlap} provides the separation performance as well as the model size and complexity for different overlap ratios between groups. We observe that adding a 25\% percent overlap between groups increases the number of MAC operations while not leading to an better performance, but a 50\% overlap between groups can improve the overall performance. Compared with the model in Table~\ref{tab:result} with a similar performance ($K=8$, $N=128$, $L_c=2$), such a model has a smaller model size and fewer MAC operations. This shows that compared with using a smaller number of groups $K$, adding proper overlap between groups is a more effective method for improving the performance.

\begin{table*}[!t]
	\scriptsize
	\centering
	\caption{Comparison of DPRNN, TCN, Sudo rm -rf, and DPTNet architectures with and without GC3. The training and inference phase statistics are evaluated with a batch size of 4.}
	\begin{tabular}{c|c|c|c|c|c|c|c|c|c}
		\thline
		\multirow{2}{*}{Model} & \multirow{2}{*}{SI-SDR (dB)} & \multirow{2}{*}{PESQ} & \multirow{2}{*}{STOI} & \multirow{2}{*}{Model size} & \multirow{2}{*}{MACs} & Training & Training & Inference & Inference \\
		 & & & & & & memory (GB) & speed (ms) & memory (MB) & speed (ms) \\
		\thline
		Mixture & -0.4 & 1.91 & 0.77 & -- & -- & -- & -- & -- & -- \\
		\hline
		DPRNN \cite{luo2020dual} & 9.0 & 2.33 & 0.80 & 2.6M & 22.1G & \textbf{3.0} & \textbf{193.4} & 24.9 & 59.7 \\
		+ GC3 & \textbf{9.1} & \textbf{2.36} & \textbf{0.82} & \textbf{123.8K (4.7\%)} & \textbf{3.9G (17.6\%)} & 4.2 & 211.3 & \textbf{6.3} & \textbf{57.2} \\
		\hline
		TCN \cite{luo2019conv} & 7.1 & 2.21 & 0.80 & 2.5M & 10.3G & \textbf{3.2} & 254.6 & 14.4 & 53.8 \\
		+ GC3 & \textbf{8.9} & \textbf{2.35} & \textbf{0.82} & \textbf{191.2K (7.6\%)} & \textbf{3.4G (33.0\%)} & 3.8 & \textbf{194.1} & \textbf{6.6} & \textbf{49.3} \\
		\hline
		Sudo rm -rf \cite{tzinis2020sudo} & 6.8 & 2.15 & 0.79 & 2.4M & 9.5G & 4.6 & 234.8 & 14.4 & 53.5 \\
		+ GC3 & \textbf{8.7} & \textbf{2.34} & \textbf{0.81} & \textbf{60.0K (2.5\%)} & \textbf{3.2G (33.7\%)} & \textbf{3.7} & \textbf{200.3} & \textbf{5.3} & \textbf{47.4} \\
		\hline
		DPTNet \cite{chen2020dual} & 8.1 & 2.20 & 0.79 & 2.8M & 21.8G & \textbf{4.8} & \textbf{256.2} & 21.2 & 78.6 \\
		+ GC3 & \textbf{8.5} & \textbf{2.32} & \textbf{0.81} & \textbf{128.6K (4.6\%)} & \textbf{3.9G (17.9\%)} & \textbf{4.8} & 272.5 & \textbf{6.2} & \textbf{76.7} \\
		\thline
	\end{tabular}
	\label{tab:tasnet}
\end{table*}

\subsection{Effect of GC3 in Different Separation Modules}
\label{sec:GC3-model}

To evaluate GC3 on the three other separation modules described in Section~\ref{sec:separation-network}, we select the hyperparameters so that all four models have on-par model size when no GroupComm or context codec are applied:

\begin{enumerate}
    \item \textit{TCN}: We apply 2 TCNs with 6 convolutional blocks in each TCN. We use the same number of TCN layers and convolutional blocks for the GC3-equipped modification.
    
    \item \textit{Sudo rm -rf}: We use the default configuration, which contains 5 downsampling and upsampling layers in each U-net block, and we use 8 blocks for both baseline and GC3-equipped modification.
    
    \item \textit{DPTNet}: We use the default configuration that contains 6 Transformer layers, and for GC3-equipped DPTNet we use 8 Transformer layers similar to our configuration for GC3-DPRNN. The learning rate warm-up configuration is also set the same as the recommended configuration, where the first 4000 iterations are used for the warm-up stage.
\end{enumerate}

The other hyperparameters are kept the same as the selected best GC3-DPRNN model in Table~\ref{tab:GroupComm}. Table~\ref{tab:tasnet} presents the separation performance as well as the model size and complexity of the four architectures with their GC3-equipped modifications. We first compare the performance of the baseline models without GC3. The plain DPRNN architecture achieves the best performance among the four architectures and is even better than the plain DPTNet, which indicates that transformer-based architectures might not be always superior than recurrent neural networks. TCN does not have satisfying performance because of the limited receptive field in our configuration (253 frames or 0.253s), as it has been shown that large receptive fields for TCN lead to better separation performance \cite{luo2019conv}. Although the selected Sudo rm -rf configuration has a large enough receptive field to cover the entire sequential feature, it obtains an even worse separation performance with on-par model size and complexity as the TCN architecture. Although \cite{tzinis2020sudo} reported that the Sudo rm -rf architecture achieved constantly better performance than DPRNN and TCN architectures, the results here indicates that its performance on the more challenging noisy reverberant environments needs to be revised. Moreover, although all four architectures achieve significant SI-SDR improvement with respect to the unprocessed mixture, the improvement on wideband PESQ and STOI scores are moderate. One possible reason for this phenomenon is the inconsistency between the time-domain and frequency-domain evaluation metrics \cite{luo2019conv}, as all the models are trained with the time-domain objective (negative SNR) while the calculation of PESQ and STOI are both in frequency domain. 

We then compare the separation performance and the model size and complexity on the GC3-equipped architectures. For DPRNN and DPTNet, GC3-equipped modifications can achieve a same level of separation performance with significantly smaller model sizes and number of MAC operations. For CNN-based architectures (TCN and Sudo rm -rf), GC3-equipped modifications can further achieve significantly higher SI-SDR scores. Since the context codec squeezes the long sequence by a factor of $C/2$ (16 for $C=32$), the effective temporal receptive field of the TCN separator is significantly larger ($0.253\times16=4.05\text{s}$) and thus can better capture the temporal dependencies. Since it has also been reported in \cite{tzinis2020sudo} that a deeper Sudo rm -rf architecture can lead to better overall separation performance, introducing GC3 to Sudo rm -rf might also be equivalent to increasing the model depth and improves the performance. More in-depth analysis on the reason behind the performance improvements in different architectures is left for future work. Nevertheless, the results prove that GC3 can be easily deployed in to various architectures and maintain its effectiveness.

Beyond the model size and number of MAC operations, the memory footprint and the training and inference speed are also important indicators for model complexity, as small models can also be slow and require enormous memory. To compare such training and inference phase statistics of different models, we evaluate the batch-level training and inference phase memory footprints and running speeds on a single NVIDIA TITAN X Pascal graphic card with a batch size of 4. The memory footprint is calculated via an opensource toolbox \cite{pytorch_memlab}. We observe that the GC3-equipped models, e.g. DPRNN and TCN, increase the training phase memory footprint but do not affect the training speed, and an acceleration can even be observed in the GC3-equipped TCN model. The inference phase memory footprint for all four architectures with GC3 applied are significantly lower than the ones without GC3, however the inference speed are all on-par or only slightly faster than the baselines. The reason for this might be because the effective model depth for a GC3-equipped model is larger than the baseline which prevents the model from easy parallelization. The results show that although the number of intermediate outputs introduced by the deeper separation module and the GroupComm modules may increase the training phase memory footprint, GC3 can always decrease the inference phase memory footprint without sacrificing the inference speed. 

\section{Conclusion and future works}
\label{sec:conclusion}
In this paper, we proposed \textit{\textbf{G}}roup \textit{\textbf{C}}ommunication with \textit{\textbf{C}}ontext \textit{\textbf{C}}odec (\textit{\textbf{GC3}}), a simple module for significantly decreasing both the model size and complexity while maintaining the performance and inference speed in various speech separation network architectures. The group communication (GroupComm) module in GC3 splits a high-dimensional feature into groups of low-dimensional features and applies a module to capture the inter-group dependency. Instead of concatenating the low-dimensional features back to a high-dimensional feature for further processing, a model with a significantly smaller width is shared by all groups and run in parallel. A context codec is applied to decrease the length of a sequential feature and thus decrease the complexity of the sequence modeling module. A context encoder compresses the temporal context of local features into a single feature, and a context decoder decompresses the transformed feature back to the context features. We conducted experiments on four different types of separation module architectures and showed that adding GC3 to the baselines led to on-par or better separation performance with significantly smaller model sizes and complexity without sacrificing the inference speed, proving that GC3 can be easily deployed to a wide range of architectures to change them into their lightweight counterparts.

For future works, we would like to investigate three important and interesting topics. The first topic is about the application of GC3 in other tasks beyond speech separation, which is a straightforward extension of the experiment configurations presented here. The second topic is to apply model binarization and quantization techniques together with GC3 to create models with even smaller model size and complexity. The third topic is to use GC3 as a prototype module and utilize neural architecture search (NAS) algorithms to search for model architectures and module organizations that better balance the model size, complexity, speed and performance.

\section{Acknowledgments}
This work was funded by a grant from the National Institute of Health, NIDCD, DC014279; and a grant from Marie-Josée and Henry R. Kravis.

\bibliographystyle{IEEEbib}
\bibliography{refs}

\begin{thebibliography}{10}
\providecommand{\url}[1]{#1}
\csname url@samestyle\endcsname
\providecommand{\newblock}{\relax}
\providecommand{\bibinfo}[2]{#2}
\providecommand{\BIBentrySTDinterwordspacing}{\spaceskip=0pt\relax}
\providecommand{\BIBentryALTinterwordstretchfactor}{4}
\providecommand{\BIBentryALTinterwordspacing}{\spaceskip=\fontdimen2\font plus
\BIBentryALTinterwordstretchfactor\fontdimen3\font minus
  \fontdimen4\font\relax}
\providecommand{\BIBforeignlanguage}[2]{{%
\expandafter\ifx\csname l@#1\endcsname\relax
\typeout{** WARNING: IEEEtran.bst: No hyphenation pattern has been}%
\typeout{** loaded for the language `#1'. Using the pattern for}%
\typeout{** the default language instead.}%
\else
\language=\csname l@#1\endcsname
\fi
#2}}
\providecommand{\BIBdecl}{\relax}
\BIBdecl

\bibitem{yu2017permutation}
D.~Yu, M.~Kolb{\ae}k, Z.-H. Tan, and J.~Jensen, ``Permutation invariant
  training of deep models for speaker-independent multi-talker speech
  separation,'' in \emph{Acoustics, Speech and Signal Processing (ICASSP), 2017
  IEEE International Conference on}.\hskip 1em plus 0.5em minus 0.4em\relax
  IEEE, 2017, pp. 241--245.

\bibitem{kolbaek2017multitalker}
M.~Kolb{\ae}k, D.~Yu, Z.-H. Tan, and J.~Jensen, ``Multitalker speech separation
  with utterance-level permutation invariant training of deep recurrent neural
  networks,'' \emph{IEEE/ACM Transactions on Audio, Speech, and Language
  Processing (TASLP)}, vol.~25, no.~10, pp. 1901--1913, 2017.

\bibitem{isik2016single}
Y.~Isik, J.~Le~Roux, Z.~Chen, S.~Watanabe, and J.~R. Hershey, ``Single-channel
  multi-speaker separation using deep clustering,'' \emph{Proc. Interspeech},
  pp. 545--549, 2016.

\bibitem{luo2017speaker}
\BIBentryALTinterwordspacing
Y.~Luo, Z.~Chen, and N.~Mesgarani, ``Speaker-independent speech separation with
  deep attractor network,'' \emph{IEEE/ACM Transactions on Audio, Speech, and
  Language Processing (TASLP)}, vol.~26, no.~4, pp. 787--796, 2018. [Online].
  Available: \url{http://dx.doi.org/10.1109/TASLP.2018.2795749}
\BIBentrySTDinterwordspacing

\bibitem{chandna2017monoaural}
P.~Chandna, M.~Miron, J.~Janer, and E.~G{\'o}mez, ``Monoaural audio source
  separation using deep convolutional neural networks,'' in \emph{International
  conference on latent variable analysis and signal separation}.\hskip 1em plus
  0.5em minus 0.4em\relax Springer, 2017, pp. 258--266.

\bibitem{takahashi2017multi}
N.~Takahashi and Y.~Mitsufuji, ``Multi-scale multi-band {D}ense{N}ets for audio
  source separation,'' in \emph{2017 IEEE Workshop on Applications of Signal
  Processing to Audio and Acoustics (WASPAA)}.\hskip 1em plus 0.5em minus
  0.4em\relax IEEE, 2017, pp. 21--25.

\bibitem{yin2020phasen}
D.~Yin, C.~Luo, Z.~Xiong, and W.~Zeng, ``{PHASEN}: A phase-and-harmonics-aware
  speech enhancement network.'' in \emph{AAAI}, 2020, pp. 9458--9465.

\bibitem{takahashi2020d3net}
N.~Takahashi and Y.~Mitsufuji, ``D3{N}et: Densely connected multidilated
  {D}ense{N}et for music source separation,'' \emph{arXiv preprint
  arXiv:2010.01733}, 2020.

\bibitem{stoller2018wave}
D.~Stoller, S.~Ewert, and S.~Dixon, ``Wave-{U}-{N}et: A multi-scale neural
  network for end-to-end audio source separation.'' in \emph{ISMIR}, 2018, pp.
  334--340.

\bibitem{venkataramani2018end}
S.~Venkataramani, J.~Casebeer, and P.~Smaragdis, ``End-to-end source separation
  with adaptive front-ends,'' in \emph{2018 52nd Asilomar Conference on
  Signals, Systems, and Computers}.\hskip 1em plus 0.5em minus 0.4em\relax
  IEEE, 2018, pp. 684--688.

\bibitem{luo2019conv}
Y.~Luo and N.~Mesgarani, ``Conv-{T}as{N}et: Surpassing ideal time--frequency
  magnitude masking for speech separation,'' \emph{IEEE/ACM Transactions on
  Audio, Speech, and Language Processing (TASLP)}, vol.~27, no.~8, pp.
  1256--1266, 2019.

\bibitem{lluis2019end}
F.~Llu{\'\i}s, J.~Pons, and X.~Serra, ``End-to-end music source separation: Is
  it possible in the waveform domain?'' \emph{Proc. Interspeech}, pp.
  4619--4623, 2019.

\bibitem{pandey2019new}
A.~Pandey and D.~Wang, ``A new framework for {CNN}-based speech enhancement in
  the time domain,'' \emph{IEEE/ACM Transactions on Audio, Speech, and Language
  Processing (TASLP)}, vol.~27, no.~7, pp. 1179--1188, 2019.

\bibitem{zhang2020furcanext}
L.~Zhang, Z.~Shi, J.~Han, A.~Shi, and D.~Ma, ``Furca{N}e{X}t: End-to-end
  monaural speech separation with dynamic gated dilated temporal convolutional
  networks,'' in \emph{International Conference on Multimedia Modeling}.\hskip
  1em plus 0.5em minus 0.4em\relax Springer, 2020, pp. 653--665.

\bibitem{tzinis2020sudo}
E.~Tzinis, Z.~Wang, and P.~Smaragdis, ``Sudo rm-rf: Efficient networks for
  universal audio source separation,'' in \emph{2020 IEEE 30th International
  Workshop on Machine Learning for Signal Processing (MLSP)}.\hskip 1em plus
  0.5em minus 0.4em\relax IEEE, 2020, pp. 1--6.

\bibitem{zeghidour2020wavesplit}
N.~Zeghidour and D.~Grangier, ``Wavesplit: End-to-end speech separation by
  speaker clustering,'' \emph{arXiv preprint arXiv:2002.08933}, 2020.

\bibitem{sun2017multiple}
L.~Sun, J.~Du, L.-R. Dai, and C.-H. Lee, ``Multiple-target deep learning for
  {LSTM}-{RNN} based speech enhancement,'' in \emph{2017 Hands-free Speech
  Communications and Microphone Arrays (HSCMA)}.\hskip 1em plus 0.5em minus
  0.4em\relax IEEE, 2017, pp. 136--140.

\bibitem{naithani2017low}
G.~Naithani, T.~Barker, G.~Parascandolo, L.~Bramsl, N.~H. Pontoppidan,
  T.~Virtanen \emph{et~al.}, ``Low latency sound source separation using
  convolutional recurrent neural networks,'' in \emph{2017 IEEE Workshop on
  Applications of Signal Processing to Audio and Acoustics (WASPAA)}.\hskip 1em
  plus 0.5em minus 0.4em\relax IEEE, 2017, pp. 71--75.

\bibitem{zhao2018convolutional}
H.~Zhao, S.~Zarar, I.~Tashev, and C.-H. Lee, ``Convolutional-recurrent neural
  networks for speech enhancement,'' in \emph{Acoustics, Speech and Signal
  Processing (ICASSP), 2018 IEEE International Conference on}.\hskip 1em plus
  0.5em minus 0.4em\relax IEEE, 2018, pp. 2401--2405.

\bibitem{tan2018convolutional}
K.~Tan and D.~Wang, ``A convolutional recurrent neural network for real-time
  speech enhancement.'' in \emph{Proc. Interspeech}, 2018, pp. 3229--3233.

\bibitem{hu2020dccrn}
Y.~Hu, Y.~Liu, S.~Lv, M.~Xing, S.~Zhang, Y.~Fu, J.~Wu, B.~Zhang, and L.~Xie,
  ``{DCCRN}: Deep complex convolution recurrent network for phase-aware speech
  enhancement,'' \emph{arXiv preprint arXiv:2008.00264}, 2020.

\bibitem{luo2020dual}
Y.~Luo, Z.~Chen, and T.~Yoshioka, ``Dual-path {RNN}: efficient long sequence
  modeling for time-domain single-channel speech separation,'' in
  \emph{Acoustics, Speech and Signal Processing (ICASSP), 2020 IEEE
  International Conference on}.\hskip 1em plus 0.5em minus 0.4em\relax IEEE,
  2020, pp. 46--50.

\bibitem{nachmani2020voice}
E.~Nachmani, Y.~Adi, and L.~Wolf, ``Voice separation with an unknown number of
  multiple speakers,'' \emph{arXiv preprint arXiv:2003.01531}, 2020.

\bibitem{chen2020dual}
J.~Chen, Q.~Mao, and D.~Liu, ``Dual-path transformer network: Direct
  context-aware modeling for end-to-end monaural speech separation,''
  \emph{Proc. Interspeech}, pp. 2642--2646, 2020.

\bibitem{kinoshita2020multi}
K.~Kinoshita, T.~von Neumann, M.~Delcroix, T.~Nakatani, and R.~Haeb-Umbach,
  ``Multi-path {RNN} for hierarchical modeling of long sequential data and its
  application to speaker stream separation,'' \emph{Proc. Interspeech 2020},
  pp. 2652--2656, 2020.

\bibitem{mazzawi2019improving}
H.~Mazzawi, X.~Gonzalvo, A.~Kracun, P.~Sridhar, N.~Subrahmanya,
  I.~Lopez-Moreno, H.-J. Park, and P.~Violette, ``Improving keyword spotting
  and language identification via neural architecture search at scale.'' in
  \emph{Proc. Interspeech}, 2019, pp. 1278--1282.

\bibitem{hu2020neural}
S.~Hu, X.~Xie, S.~Liu, M.~Geng, X.~Liu, and H.~Meng, ``Neural architecture
  search for speech recognition,'' \emph{arXiv preprint arXiv:2007.08818},
  2020.

\bibitem{aihara2019teacher}
R.~Aihara, T.~Hanazawa, Y.~Okato, G.~Wichern, and J.~Le~Roux, ``Teacher-student
  deep clustering for low-delay single channel speech separation,'' in
  \emph{Acoustics, Speech and Signal Processing (ICASSP), 2019 IEEE
  International Conference on}.\hskip 1em plus 0.5em minus 0.4em\relax IEEE,
  2019, pp. 690--694.

\bibitem{kim2018bitwise}
M.~Kim and P.~Smaragdis, ``Bitwise neural networks for efficient single-channel
  source separation,'' in \emph{Acoustics, Speech and Signal Processing
  (ICASSP), 2018 IEEE International Conference on}.\hskip 1em plus 0.5em minus
  0.4em\relax IEEE, 2018, pp. 701--705.

\bibitem{kim2019incremental}
S.~Kim, M.~Maity, and M.~Kim, ``Incremental binarization on recurrent neural
  networks for single-channel source separation,'' in \emph{Acoustics, Speech
  and Signal Processing (ICASSP), 2019 IEEE International Conference on}.\hskip
  1em plus 0.5em minus 0.4em\relax IEEE, 2019, pp. 376--380.

\bibitem{kim2020boosted}
S.~Kim, H.~Yang, and M.~Kim, ``Boosted locality sensitive hashing:
  Discriminative binary codes for source separation,'' in \emph{Acoustics,
  Speech and Signal Processing (ICASSP), 2020 IEEE International Conference
  on}.\hskip 1em plus 0.5em minus 0.4em\relax IEEE, 2020, pp. 106--110.

\bibitem{hinton2015distilling}
G.~Hinton, O.~Vinyals, and J.~Dean, ``Distilling the knowledge in a neural
  network,'' \emph{arXiv preprint arXiv:1503.02531}, 2015.

\bibitem{luo2017thinet}
J.-H. Luo, J.~Wu, and W.~Lin, ``Thi{N}et: A filter level pruning method for
  deep neural network compression,'' in \emph{Proceedings of the IEEE
  international conference on computer vision}, 2017, pp. 5058--5066.

\bibitem{simons2019review}
T.~Simons and D.-J. Lee, ``A review of binarized neural networks,''
  \emph{Electronics}, vol.~8, no.~6, p. 661, 2019.

\bibitem{luo2020ultra}
Y.~Luo, C.~Han, and N.~Mesgarani, ``Ultra-lightweight speech separation via
  group communication,'' \emph{arXiv preprint arXiv:2011.08397}, 2020.

\bibitem{li2015lstm}
J.~Li, A.~Mohamed, G.~Zweig, and Y.~Gong, ``{LSTM} time and frequency
  recurrence for automatic speech recognition,'' in \emph{Automatic Speech
  Recognition and Understanding (ASRU), 2015 IEEE Workshop on}.\hskip 1em plus
  0.5em minus 0.4em\relax IEEE, 2015, pp. 187--191.

\bibitem{wang2017joint}
Q.~Wang, J.~Du, L.-R. Dai, and C.-H. Lee, ``Joint noise and mask aware training
  for {DNN}-based speech enhancement with sub-band features,'' in \emph{2017
  Hands-free Speech Communications and Microphone Arrays (HSCMA)}.\hskip 1em
  plus 0.5em minus 0.4em\relax IEEE, 2017, pp. 101--105.

\bibitem{zhang2017deep}
X.~Zhang and D.~Wang, ``Deep learning based binaural speech separation in
  reverberant environments,'' \emph{IEEE/ACM Transactions on Audio, Speech, and
  Language Processing (TASLP)}, vol.~25, no.~5, pp. 1075--1084, 2017.

\bibitem{li2019multichannel}
X.~Li and R.~Horaud, ``Multichannel speech enhancement based on time-frequency
  masking using subband long short-term memory,'' in \emph{2019 IEEE Workshop
  on Applications of Signal Processing to Audio and Acoustics (WASPAA)}.\hskip
  1em plus 0.5em minus 0.4em\relax IEEE, 2019, pp. 298--302.

\bibitem{li2016exploring}
J.~Li, A.~Mohamed, G.~Zweig, and Y.~Gong, ``Exploring multidimensional {LSTM}s
  for large vocabulary {ASR},'' in \emph{Acoustics, Speech and Signal
  Processing (ICASSP), 2016 IEEE International Conference on}.\hskip 1em plus
  0.5em minus 0.4em\relax IEEE, 2016, pp. 4940--4944.

\bibitem{sainath2016modeling}
T.~N. Sainath and B.~Li, ``Modeling time-frequency patterns with {LSTM} vs.
  convolutional architectures for lvcsr tasks,'' \emph{Proc. Interspeech}, pp.
  813--817, 2016.

\bibitem{xu2018single}
C.~Xu, W.~Rao, X.~Xiao, E.~S. Chng, and H.~Li, ``Single channel speech
  separation with constrained utterance level permutation invariant training
  using grid {LSTM},'' in \emph{Acoustics, Speech and Signal Processing
  (ICASSP), 2018 IEEE International Conference on}.\hskip 1em plus 0.5em minus
  0.4em\relax IEEE, 2018, pp. 6--10.

\bibitem{whitehead2011precision}
N.~Whitehead and A.~Fit-Florea, ``Precision \& performance: Floating point and
  {IEEE} 754 compliance for {NVIDIA} {GPU}s,'' \emph{rn (A+ B)}, vol.~21,
  no.~1, pp. 18\,749--19\,424, 2011.

\bibitem{luo2018tasnet}
Y.~Luo and N.~Mesgarani, ``Tas{N}et: time-domain audio separation network for
  real-time, single-channel speech separation,'' in \emph{Acoustics, Speech and
  Signal Processing (ICASSP), 2018 IEEE International Conference on}.\hskip 1em
  plus 0.5em minus 0.4em\relax IEEE, 2018.

\bibitem{hochreiter1997long}
S.~Hochreiter and J.~Schmidhuber, ``Long short-term memory,'' \emph{Neural
  computation}, vol.~9, no.~8, pp. 1735--1780, 1997.

\bibitem{ba2016layer}
J.~L. Ba, J.~R. Kiros, and G.~E. Hinton, ``Layer normalization,'' \emph{arXiv
  preprint arXiv:1607.06450}, 2016.

\bibitem{gao2019res2net}
S.~Gao, M.-M. Cheng, K.~Zhao, X.-Y. Zhang, M.-H. Yang, and P.~H. Torr,
  ``Res2{N}et: A new multi-scale backbone architecture,'' \emph{IEEE
  transactions on pattern analysis and machine intelligence (TPAMI)}, 2019.

\bibitem{zhang2020resnest}
H.~Zhang, C.~Wu, Z.~Zhang, Y.~Zhu, Z.~Zhang, H.~Lin, Y.~Sun, T.~He, J.~Mueller,
  R.~Manmatha \emph{et~al.}, ``Res{N}e{S}t: Split-attention networks,''
  \emph{arXiv preprint arXiv:2004.08955}, 2020.

\bibitem{han2020ghostnet}
K.~Han, Y.~Wang, Q.~Tian, J.~Guo, C.~Xu, and C.~Xu, ``Ghost{N}et: More features
  from cheap operations,'' in \emph{Proceedings of the IEEE/CVF Conference on
  Computer Vision and Pattern Recognition}, 2020, pp. 1580--1589.

\bibitem{xu2014regression}
Y.~Xu, J.~Du, L.-R. Dai, and C.-H. Lee, ``A regression approach to speech
  enhancement based on deep neural networks,'' \emph{IEEE/ACM Transactions on
  Audio, Speech, and Language Processing (TASLP)}, vol.~23, no.~1, pp. 7--19,
  2014.

\bibitem{araki2015exploring}
S.~Araki, T.~Hayashi, M.~Delcroix, M.~Fujimoto, K.~Takeda, and T.~Nakatani,
  ``Exploring multi-channel features for denoising-autoencoder-based speech
  enhancement,'' in \emph{Acoustics, Speech and Signal Processing (ICASSP),
  2015 IEEE International Conference on}.\hskip 1em plus 0.5em minus
  0.4em\relax IEEE, 2015, pp. 116--120.

\bibitem{chen2015long}
X.~Chen, X.~Qiu, C.~Zhu, P.~Liu, and X.-J. Huang, ``Long short-term memory
  neural networks for chinese word segmentation,'' in \emph{Proceedings of the
  2015 Conference on Empirical Methods in Natural Language Processing}, 2015,
  pp. 1197--1206.

\bibitem{pathak2016context}
D.~Pathak, P.~Krahenbuhl, J.~Donahue, T.~Darrell, and A.~A. Efros, ``Context
  encoders: Feature learning by inpainting,'' in \emph{Proceedings of the IEEE
  conference on computer vision and pattern recognition}, 2016, pp. 2536--2544.

\bibitem{mehri2016samplernn}
S.~Mehri, K.~Kumar, I.~Gulrajani, R.~Kumar, S.~Jain, J.~Sotelo, A.~Courville,
  and Y.~Bengio, ``Sample{RNN}: An unconditional end-to-end neural audio
  generation model,'' \emph{arXiv preprint arXiv:1612.07837}, 2016.

\bibitem{marafioti2019context}
A.~Marafioti, N.~Perraudin, N.~Holighaus, and P.~Majdak, ``A context encoder
  for audio inpainting,'' \emph{IEEE/ACM Transactions on Audio, Speech, and
  Language Processing (TASLP)}, vol.~27, no.~12, pp. 2362--2372, 2019.

\bibitem{luo2020implicit}
Y.~Luo and N.~Mesgarani, ``Implicit filter-and-sum network for multi-channel
  speech separation,'' \emph{arXiv preprint arXiv:2011.08401}, 2020.

\bibitem{luo2020end}
Y.~Luo, Z.~Chen, N.~Mesgarani, and T.~Yoshioka, ``End-to-end microphone
  permutation and number invariant multi-channel speech separation,'' in
  \emph{Acoustics, Speech and Signal Processing (ICASSP), 2020 IEEE
  International Conference on}.\hskip 1em plus 0.5em minus 0.4em\relax IEEE,
  2020, pp. 6394--6398.

\bibitem{panayotov2015librispeech}
V.~Panayotov, G.~Chen, D.~Povey, and S.~Khudanpur, ``Librispeech: an {ASR}
  corpus based on public domain audio books,'' in \emph{Acoustics, Speech and
  Signal Processing (ICASSP), 2015 IEEE International Conference on}.\hskip 1em
  plus 0.5em minus 0.4em\relax IEEE, 2015, pp. 5206--5210.

\bibitem{web100nonspeech}
G.~Hu, ``100 {N}onspeech {S}ounds,'' {\small
  \url{http://web.cse.ohio-state.edu/pnl/corpus/HuNonspeech/HuCorpus.html}}.

\bibitem{allen1979image}
J.~B. Allen and D.~A. Berkley, ``Image method for efficiently simulating
  small-room acoustics,'' \emph{The Journal of the Acoustical Society of
  America}, vol.~65, no.~4, pp. 943--950, 1979.

\bibitem{diaz2020gpurir}
D.~Diaz-Guerra, A.~Miguel, and J.~R. Beltran, ``gpu{RIR}: A python library for
  room impulse response simulation with gpu acceleration,'' \emph{Multimedia
  Tools and Applications}, pp. 1--19, 2020.

\bibitem{he2015delving}
K.~He, X.~Zhang, S.~Ren, and J.~Sun, ``Delving deep into rectifiers: Surpassing
  human-level performance on imagenet classification,'' in \emph{Proceedings of
  the IEEE international conference on computer vision}, 2015, pp. 1026--1034.

\bibitem{vaswani2017attention}
A.~Vaswani, N.~Shazeer, N.~Parmar, J.~Uszkoreit, L.~Jones, A.~N. Gomez,
  {\L}.~Kaiser, and I.~Polosukhin, ``Attention is all you need,'' in
  \emph{Advances in neural information processing systems}, 2017, pp.
  5998--6008.

\bibitem{dong2018speech}
L.~Dong, S.~Xu, and B.~Xu, ``Speech-transformer: a no-recurrence
  sequence-to-sequence model for speech recognition,'' in \emph{Acoustics,
  Speech and Signal Processing (ICASSP), 2018 IEEE International Conference
  on}.\hskip 1em plus 0.5em minus 0.4em\relax IEEE, 2018, pp. 5884--5888.

\bibitem{li2019neural}
N.~Li, S.~Liu, Y.~Liu, S.~Zhao, and M.~Liu, ``Neural speech synthesis with
  transformer network,'' in \emph{Proceedings of the AAAI Conference on
  Artificial Intelligence}, vol.~33, 2019, pp. 6706--6713.

\bibitem{tsunoo2019transformer}
E.~Tsunoo, Y.~Kashiwagi, T.~Kumakura, and S.~Watanabe, ``Transformer {ASR} with
  contextual block processing,'' in \emph{Automatic Speech Recognition and
  Understanding (ASRU), 2019 IEEE Workshop on}.\hskip 1em plus 0.5em minus
  0.4em\relax IEEE, 2019, pp. 427--433.

\bibitem{karita2019comparative}
S.~Karita, N.~Chen, T.~Hayashi, T.~Hori, H.~Inaguma, Z.~Jiang, M.~Someki,
  N.~E.~Y. Soplin, R.~Yamamoto, X.~Wang \emph{et~al.}, ``A comparative study on
  transformer vs {RNN} in speech applications,'' in \emph{Automatic Speech
  Recognition and Understanding (ASRU), 2019 IEEE Workshop on}.\hskip 1em plus
  0.5em minus 0.4em\relax IEEE, 2019, pp. 449--456.

\bibitem{chollet2016xception}
F.~Chollet, ``Xception: Deep learning with depthwise separable convolutions,''
  \emph{arXiv preprint}, 2016.

\bibitem{luo2020distortion}
Y.~Luo, C.~Han, and N.~Mesgarani, ``Distortion-controlled training for
  end-to-end reverberant speech separation with auxiliary autoencoding loss,''
  in \emph{2021 IEEE Spoken Language Technology Workshop (SLT)}.\hskip 1em plus
  0.5em minus 0.4em\relax IEEE, 2021.

\bibitem{le2019sdr}
J.~Le~Roux, S.~Wisdom, H.~Erdogan, and J.~R. Hershey, ``{SDR}--half-baked or
  well done?'' in \emph{Acoustics, Speech and Signal Processing (ICASSP), 2019
  IEEE International Conference on}.\hskip 1em plus 0.5em minus 0.4em\relax
  IEEE, 2019, pp. 626--630.

\bibitem{rix2001perceptual}
A.~W. Rix, J.~G. Beerends, M.~P. Hollier, and A.~P. Hekstra, ``Perceptual
  evaluation of speech quality ({PESQ}) - a new method for speech quality
  assessment of telephone networks and codecs,'' in \emph{Acoustics, Speech and
  Signal Processing (ICASSP), 2001 IEEE International Conference on},
  vol.~2.\hskip 1em plus 0.5em minus 0.4em\relax IEEE, 2001, pp. 749--752.

\bibitem{taal2010short}
C.~H. Taal, R.~C. Hendriks, R.~Heusdens, and J.~Jensen, ``A short-time
  objective intelligibility measure for time-frequency weighted noisy speech,''
  in \emph{Acoustics, Speech and Signal Processing (ICASSP), 2010 IEEE
  International Conference on}.\hskip 1em plus 0.5em minus 0.4em\relax IEEE,
  2010, pp. 4214--4217.

\bibitem{pytorch_memlab}
``pytorch\_memlab,'' {\small
  \url{https://github.com/Stonesjtu/pytorch_memlab}}.

\end{thebibliography}

\end{document}